# Remote-control and clustering of physical computations using the XML-RPC protocol and the open-Mosix system

Tomasz Błachowicz[*] and Michał Wieja

Department of Electron Technology, Institute of Physics, Silesian University of Technology

Krzywoustego 2, 44-100 Gliwice, POLAND

**Abstract**

The applications of the remote control of physical simulations performed in clustered computers running under an open-Mosix system are presented. Results from the simulation of a 2-dimensional ferromagnetic system of spins in the Ising scheme are provided. Basic parameters of a simulated hysteresis loop like coercivity and exchange bias due to pinning of ferromagnetic spins are given. The paper describes in physicists terminology a cost effective solution which utilizes an XML-RPC protocol (Extensible Markup Language - Remote Procedure Calling) and standard C++ and Python languages.

---

[*] corresponding author: Tel.: +48 32 2372071; Fax: +48 32 2372057.

E-mail address: Tomasz.Blachowicz@polsl.pl

# 1. Introduction

A typical numerical task in computational physics requires repetition of a single simulation when the simulated physical phenomenon has a statistical nature. In practice, while the single simulation can last some minutes on a PC-class machine, then time of computation for the whole set of such single evaluations extends proportionally to its number. For example, in the Ising-type simulations which use random-number generators to describe spin distributions and their dynamics, there is a need to average final parameters like the spin-system magnetization from $10^3$ or even more repeats. This makes computation ineffective on a single machine. Thus, there is a practical need to speed up using existing infrastructures to make the solution cost effective. One of the commonly applied solutions that fulfills these requirements is the open-Mosix management system [1] using Linux which enables the control of numerical processes migration using a classical Ethernet network of single PC's arranged in a cluster. The management and operation of this solution are well described in many on-line tutorials and manuals from the computer science perspective. There is also supporting software that provides visual information of numerical processes distribution within a cluster. However, from a computational physics point of view, where there is a need for repetition of the assumed number of repeated processes of simulated physical phenomena followed by the averaging of results from completed processes and the preparation of results for graphical presentations, the computer science point of view may be unclear for the physics community. It may also be desirable that the remote communication with distributed open-Mosix machines should make such work easier and similar to that of a single-user PC sometimes running under a different operating system, like Windows. Thus, a numerical code written in a given high-level language should be constructed in a way to enable the remote control of clustered physical phenomena simulations.

The information to be provided here is intended for physicists with a weak computer science background. In addition, the need of computer engineers working within computational physics will be addressed as we describe modifications within the numerical code to manage multiprocessing and remote-control performed by a single-user apart from the cluster. Thus, from the physical point of view, some issues related to the 2-dimensional Random Field Ising Model (RFIM) of ferromagnetic spins and their hysteresis loop will be given [2, 3]. Next, from a computational point of view, some examples related to C++ and Python languages will be provided. The key issue however for the numerical solution presented here is the remote control of simulations utilizing the XML-RPC technology, which is system independent.

## 2. Remote controlling of simulations using the XML-RPC protocol

The XML-RPC (Extensible Markup Language - Remote Procedure Calling) protocol enables control of programs running on remote computers [4]. The functions and procedures within a given remote application can be called using the standard HTML syntax. However, at a step prior to sending HTML, data are encoded into the XML format. Importantly, the encoded data can posses a structural character. There exist several implementations of XML-RPC depending on the language applied. These can be used with most popular C/C++ or Python languages among the many others. For example, the xmlrpclib library was added to the Python resources in 2003 [5]. Also, in the C++, the XmlRpcCpp.h module is easily available [6].

The XML-RPC protocol is a client-server type solution. In the present circumstance, the server was implemented in Python and the client was written in C++. By a client we mean the C++ code which describes the simulated phenomena. By a server we understand the software which controls the repetition of single simulations. Especially, it retains single process results, averages, and prepares these results in a column format useful for typical graphical presentations. The choice of languages results from the speed of performance; on the server side, where data aren't processed, the performance isn't a problem, so we can stay with the Python as it is an interpreted language. The advantage is that Python enables fast creation of software. On the client side the highest possible performance is required, so all the code can be written in C++. To be clear, the client and the server are the common XML-RPC terms.

Fig. 1 show schematically the XML-RPC principle of work in our solution. The right box is the graphical client application, that tells the server to start simulations and later checks the simulation progress by calling the *show_results()* function on the server side; the graphical client runs under Windows. The left box on the diagram shows a single simulation process which sends a given result to the server where it can be processed, averaged with previous results and prepared for graphical presentation in the graphical client.

Let's focus on the server implementation for a moment. We need to create a class that will contain methods accessible through XML-RPC. Obviously we can refer to this class as the *Server*:

class *Server*:
    def *start_simulations*(self, parameters_array):
        …

          return 0
     def *collect_data*(self, data):
          self.results = process_data(data)
     def *show_results*(self):
          return self.results
          return 0

This is an example of the server with the appropriate methods. In the *start_simulation* method, we usually we want to refer to the operating system method to create a certain amount of processes and the *parameters_array* should contain the assumed number of repetitions (due to statistical nature of phenomena) to run, the information of the simulated phenomena and other physical parameters that simulations should be started with. Next, the method *collect_data* is called ('called' in a meaning of XML-RPC) by a simulation processes to drop the results. This method should create data in a format ready for graphical presentation, calculate averaged results along with uncertainties. The third method *show_results* returns data suitable for the graphical client - the same array that was created by *collect_data* method.

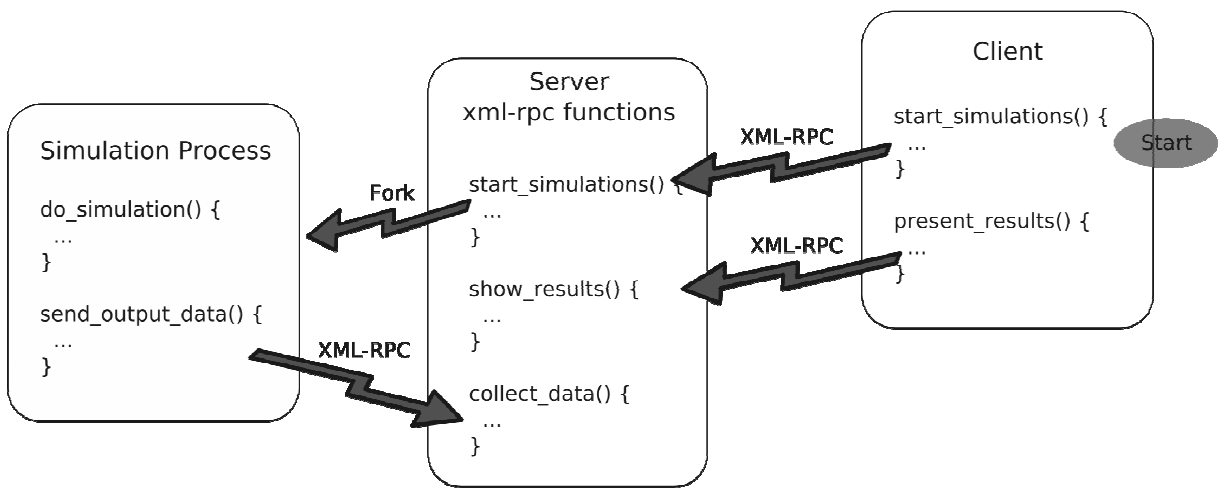

**Fig. 1.** The diagram shows the XML-RPC philosophy of performance within the client-server solution. The middle box represents a server application that does three tasks: starts a specified number of statistical simulations on a client demand, collects simulated data, and shows results of a client demand. All of these functions are available through the XML-RPC protocol - the client and simulation processes can run on separate remote machines.

Thus, our Python XML-RPC server will need to import the XML-RPC module using the *import SimpleXMLRPCServer* command. Next, we should make the *Server* class useful for a client. Thus, we have to register an object of the *Server* class as the XML-RPC server. Finally, we can start the server process to serve continually ('forever' in a meaning of XML-RPC). To do this in Python we need four lines:

```
server_object = Server()
server = SimpleXMLRPCServer.SimpleXMLRPCServer(('10.0.0.1', 8000))
server.register_instance(server_object)
server.serve_forever()
```

The first line creates an object of the class *Server*; the second one creates the XML-RPC server that will bind IP address '10.0.0.1' at the port 8000. The IP address should be our LAN-IP address accessible from the network, and the port shouldn't be already opened by any other application. The third line registers our object as an instance in the XML-RPC server, so it makes methods of this object call-able from the network. The fourth and the last line locks the XML-RPC server to be called 'forever'. This simple example explains how we can create the XML-RPC server in Python. The server should be executed on a cluster (on a Linux side) before beginning simulations using the ./server.py command, where server.py is the name of the Python script.

Now, let's take a closer look into the simulation process, which runs in a cluster. The main C++ function should contain two calls ('calls' in the XML-RPC protocol meaning). The first one collects parameters from a command line (Windows application) and passes these to our simulation. Next, the *do_simulation( )* function is called to start the clustered calculations (comp. left box in Fig. 1). The *do_simulation( )* function produces the array named 'data'. We need to send these data to the server utilizing the XML-RPC call. This is done by the function *send_output_data( )*. In C++ we need to use an external library to have access to this functionality. To use this library we need to include following header file in the main.cpp file: #include <XmlRpcCpp.h>. Then we should define the server address, for example:

```
#define SERVER_URL "http://10.0.0.1:8000"
```

and the two strings necessary to initialize xmlrpc client, namely:

```
#define NAME        "XML-RPC C++ simulation client"
#define VERSION     "0.1"
```

In the next step we can initialize the xmlrpc client using the *sent_output_data( )* function in the following style:

```
XmlRpcClient::Initialize(NAME, VERSION);
XmlRpcClient server (SERVER_URL);
```

Numerically, the xmlrpc client expects to work with the XmlRpcValue type of data. Thus, we need to copy-over numerical data to the XmlRpcValue data type. To do that we must first create the XmlRpcValue array using following command:

```
XmlRpcValue sent_table = XmlRpcValue::makeArray();
```

Then we can copy the simulated results from the data[i] array to the XmlRpc sent_table structure, as follows:

```
for (int i=0; i < number_of_elements; i++){
    sent_table.arrayAppendItem(XmlRpcValue::makeDouble(data[i]));
}
```

Finally, we send data from the clustered process back to the server using following XML-RPC command:

```
XmlRpcValue result1 = server.call("collect_data ", sent_array);
```

Now, it may be helpful to know how the above described scheme fits into the open-Mosix system functionality. Thus, Fig. 2 updates information from Fig. 1 where earlier only the XML-RPC principle of work was emphasized. First of all, using the Mosix-type solution we can make use of automatic process migrations between machines. Importantly, the migration over a network to a machine with the lowest load is automatic. The only step that a user has to take into account is the creation of several simulation processes ($10^3$) followed by the collection and averaging of resultant data from different, identified processes.

In order to achieve that, processes have to identify themselves at the moment when simulations are created ("forked" in the Unix language). This is why we assign them in the program code a process identification number ('pid'). Then, each separate process can return data with the information that this particular process has been finished. Thus, each simulation process, when sending data back by calling the *collect_data* function, identifies itself with its 'pid'. In other words, the 'pid' is a separate server function parameter, next to the simulated result array. After implementation of this kind of functionality the server knows how many processes have been created, how many processes have been finished, and how many processes are in progress.

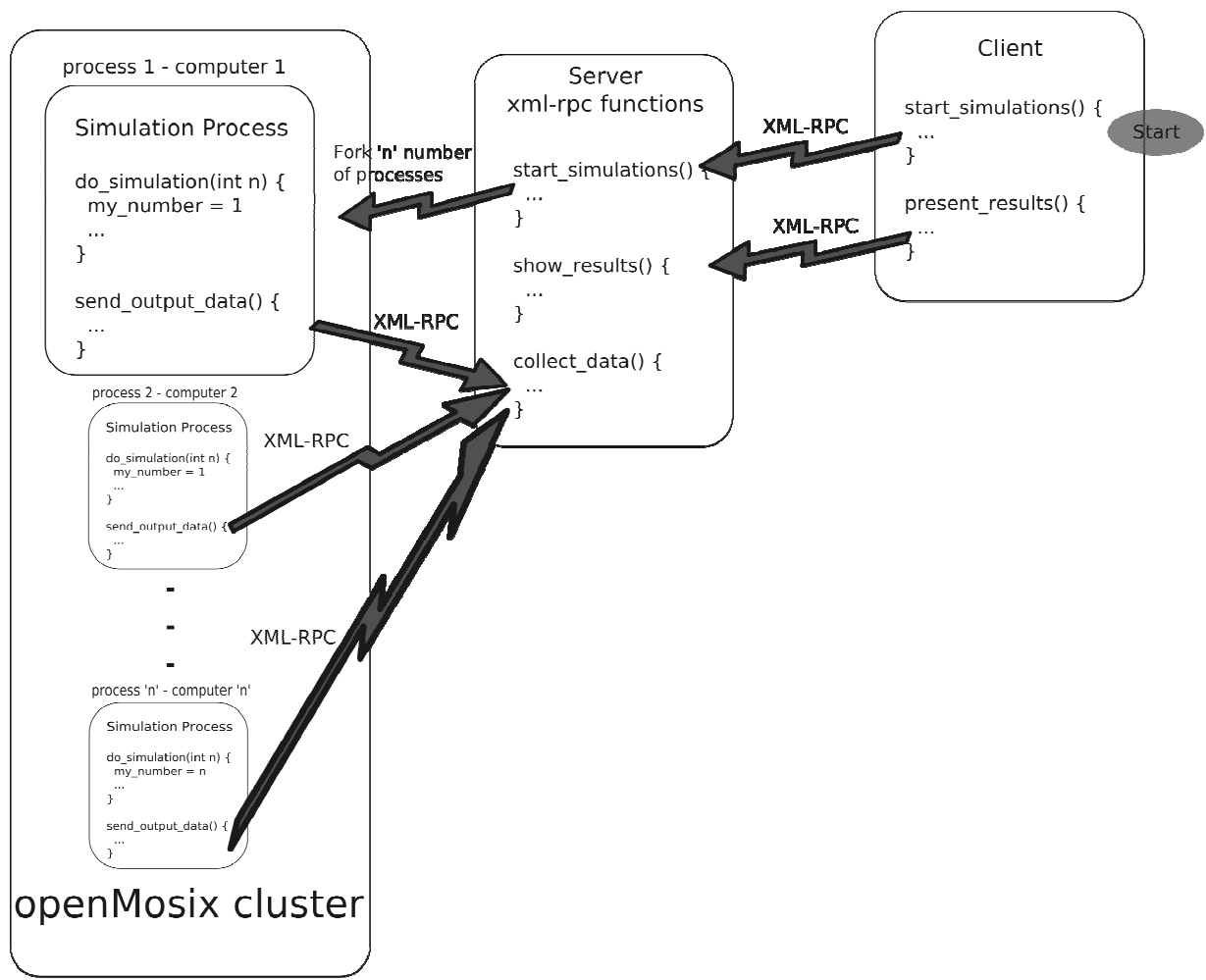

Fig. 2. The open-Mosix functionality implemented with the XML-RPC technology.

## 3. Discussing the solution efficiency applied in Ising-type simulations

In order to test the solution described in this paper, we used the Random Field Ising Model (RFIM) on the 2-dimensional 70 x 70 structure of spins. Details of the physical model applied have been described in Ref. [2, 7-9]. Within this approach, which is typical in the Ising-type calculations, the spin dynamics with the two +1 and -1 states were tested. The local condition of a spin flip from one state into another is fulfilled, when the numerical value of the total magnetic field acting on this spin changes its algebraic sign (from plus to minus, and inversely, from minus to plus). The total magnetic field $H_{tot}(i, j)$ influencing a given spin at the *(i, j)* position consists of the three following components: the exchange interaction with the four nearest neighbors (a), the random field of the Gaussian distribution *GRF(i, j)* representing natural atomic roughness of the structure, and the externally applied magnetic field $H_{ext}$ (c). The $H_{tot}(i, j)$ field can be written as follows

$$H_{tot}(i, j) = J(i, j-1) \cdot S(i, j-1) + J(i, j+1) \cdot S(i, j+1) + J(i-1, j) \cdot S(i-1, j) + J(i+1, j) \cdot S(i+1, j) + GRF(i, j) + H_{ext},$$

(1)

where $J(i, j)$ is the exchange constant and $S(i, j)$ is the spin value (+1 or -1). Additionally, at randomly distributed positions the exchange constant has been enhanced several times from its normal value of unity to block spin reversal at these locations. The reason for that pinning is that some of the spin are biased – this is typical behavior of the so-called exchange-bias phenomenon [10-12]. This phenomenon is characterized by a hysteresis loop shift of the ferromagnetic layer along the magnetic field axis due to influence of attached antiferromagnetic domains which can block the ferromagnetic spin reversal.

What we simulated here is the hysteresis loop of the (*n* x *n*) structure (*n*=70), thus the dependence between magnetization calculated as

$$M = \frac{\sum_{i,j} S(i, j)}{n^2}$$

(2)

and the externally applied magnetic field. Simulations began from the saturation of magnetization (*M*=1), with the standard deviation of the gaussially distributed roughness equal to 1.5, and the exchange constant maximized 10 times at 10% randomly distributed locations (uniform distribution). The resultant values of exchange-bias and coercivity fields were averaged from *N*=8, 16, 32, 64, 128, 512 and 1024 repetitions. The exchange bias shift was calculated as the half difference between the points where the loop crosses the field axis (abscissa), while the coercivity was obtained as the distance between these points. The calculations were carried out on the computer cluster built from 16 PC-type computers each equipped with the 512MB of RAM and the x86 processors. An Ethernet network at the 100Mb/sec. speed connected clustered computers. Table I provides the times of computation, the values of calculated exchange-bias and coercivities along with their standard deviations.

Table I. Simulated exchange-bias and coercivity field from different numbers of repetitions in the RFIM simulation of the 2-dimensional (70 x 70) spin structure.

| Number of repetitions | Exchange-bias (a. u.) | Coercivity field (a. u.) | Time of computation (sec.) |
|---|---|---|---|
| 8 | 0.380±0.011 | 4.236±0.030 | 108 |
| 16 | 0.418±0.021 | 4.225±0.015 | 210 |
| 32 | 0.419±0.009 | 4.222±0.017 | 325 |
| 64 | 0.427±0.009 | 4.228±0.014 | 580 |
| 128 | 0.419±0.006 | 4.242±0.009 | 940 |
| 256 | 0.412±0.007 | 4.186±0.031 | 1960 |
| 512 | 0.397±0.007 | 4.169±0.025 | 3552 |
| 1024 | 0.415±0.003 | 4.208±0.011 | 7175 |

The results possess a statistical character, while values of uncertainties have a tendency to minimize. The time of computation is more or less proportional to a number of repetitions. Some deviations from an exact line dependency are caused by network performance.

The simulations were controlled in a local network as well as from localizations of 1000km away. Obviously, geographical distance doesn't matter in this application. The solution presented here are extremely useful due to their remote character as well as to their independence from the operating system applied.

**Acknowledgements**

This work was supported by the Polish State Committee for Scientific Research under Grant No 3 T11F 015 26.

**References**

[1] D. Robbins, Introduction to OpenMosix, http://www.intel.com/cd/ids/developer/asmona/eng/20449.htm. More information available at www.openmosix.org.

[2] T. Błachowicz, Central European Journal of Physics, 3 (2005) 147-162.

[3] R. Blossey, T. Kinoshita, J. Dupont-Roc, Physica A, 248 (1998) 247-272.

[4] D. Winer, XML-RPC for Newbies, available on-line at: http://davenet.scripting.com/1998/07/14/xmlRpcForNewbies. The main source of information can be found at http://www.xmlrpc.com/ and related links.

[5] G. van Rossum, Python Tutorial, available on-line at http://docs.python.org/tut/tut.html.

[6] The xmlrpc library is accessible at http://xmlrpc-c.sourceforge.net.

[7] E. S. de Sousa, A. M. Mariz, F. D. Nobre, U. M. S. Costa, Physica A, 241 (1997) 469-480.

[8] U. Nowak, K. D. Usadel, J. Esser, Physica A, 250 (1998) 1-7.

[9] X. Illa, E. Vives, A. Planes, Phys. Rev. B, 66 (2002) 224422-224429.

[10] J. Nogués, I. K. Schuller, J. Mag. Mag. Mat., 192 (1999) 203-232.

[11] A. E. Berkowitz, K. Takano, J. Mag. Mag. Mat., 200 (1999) 552-570.

[12] B. Beschoten, A. Tillmanns, J. Keller, G. Güntherodt, U. Nowak, K. D. Usadel, Adv. Sol. Stat. Phys., 42 (2002) 419-431.

**Appendix A. The part of C++ code realizing the XMLRPC protocol (main.cpp)**

```
#include <XmlRpcCpp.h>

#include <iostream>
#include <stdlib.h>

#define SIZE 70;    //the structure size
#define STEPS 300;  //number of steps of magnetic field evolution from max
                    //to min values
#define HMAX 8;
#define HMIN -5;
#define DLEVEL 0.10; //dilution level = 10%
#define ECONST 10;   //enhanced exchange constant
#define SD 1.5;      //standard deviation of roughness
#define PID 0;
#define NAME       "XML-RPC getSumAndDifference C++ Client"
#define VERSION    "0.1"
#define SERVER_URL "http://10.0.0.1:8000"

...

int finish(int process_id, float _data[][2], float res[2], int
number_of_steps){

    XmlRpcClient::Initialize(NAME, VERSION);
    XmlRpcClient server (SERVER_URL);

// 'a' contains information about an external field H:

    XmlRpcValue a = XmlRpcValue::makeArray();

// 'b' contains information about the magnetization M,
//while 'a' and 'b' are used for the graphical presentation M=M(H):

    XmlRpcValue b = XmlRpcValue::makeArray();

// 'c' keeps results of a hysteresis loop shift (exchange-bias) and
// a coercivity:

    XmlRpcValue c = XmlRpcValue::makeArray();

    for (int i=0; i < 2; i++){
        c.arrayAppendItem(XmlRpcValue::makeDouble(res[i]));
    }

    for (int i=0; i < number_of_steps; i++){
        a.arrayAppendItem(XmlRpcValue::makeDouble(_data[i][0])); //H values
        b.arrayAppendItem(XmlRpcValue::makeDouble(_data[i][1])); //M values
    }

    XmlRpcValue param_array1 = XmlRpcValue::makeArray();
    param_array1.arrayAppendItem(XmlRpcValue::makeInt(process_id));
    param_array1.arrayAppendItem(a);
    XmlRpcValue result1 = server.call("store_array1",param_array1);

    XmlRpcValue param_array2 = XmlRpcValue::makeArray();
    param_array2.arrayAppendItem(XmlRpcValue::makeInt(process_id));
    param_array2.arrayAppendItem(b);
    XmlRpcValue result2 = server.call("store_array2",param_array2);
```

```
        XmlRpcValue param_array3 = XmlRpcValue::makeArray();
        param_array3.arrayAppendItem(XmlRpcValue::makeInt(process_id));
        param_array3.arrayAppendItem(c);
        XmlRpcValue result3 = server.call("store_results",param_array3);

        XmlRpcValue param_array4 = XmlRpcValue::makeArray();
        param_array4.arrayAppendItem(XmlRpcValue::makeInt(process_id));
        XmlRpcValue result4 = server.call("finish",param_array4);
}

int main(int argc, char *argv[])
{
        char *file_name;
        int size =SIZE;
        int steps = STEPS;
        float hmax = HMAX;
        float hmin = HMIN;
        float dlevel = DLEVEL;
        int econst = ECONST;
        float sd = SD;
        int pid = PID;

...

//this function, declared below, is defined in the main.cpp file:

calculate(pid, size, steps, hmax, hmin, dlevel, econst, sd, file_name);

}
```

**Appendix B. The code of the C++ calculate function realizing the simulation process**

```
#include <iostream>
#include <stdlib.h>
#include <sys/times.h>
#include <math.h>
#include <gsl/gsl_rng.h>
#include <gsl/gsl_randist.h>

...
//Here below is the beginning of definition of the calculate function.
//This should be written in a classical C++ syntax:

int calculate(int pid, int size_of_structure, int number_of_steps, float
Hmax, float Hmin, float dilution_level, int enhanced_exchange_constant,
float sd, char *fname) {

...

// This array below contains a pair of M=M(H) data:

float results[2];

...

//This function, declared below, is defined in the server.py file
//Appendix C):

  finish(pid, _data, results, number_of_steps);

  return 0;
}
```

## Appendix C. The Python code of the server.py implementation utilizing the XMPRPC protocol

```
#This tells UNIX shell that the Python interpreter has to be started to
#execute the server.py file:

#!/usr/bin/python

#Following lines adds several libraries needed by script to execute. The
#Scipy and the numarray are numerical external modules, queue is our
#internal module that is implemented in a separate file (queue.py,
#Appendix D). The thread, the SimpleXMLRPCServer, the csv, and the os are
#standard python library modules:

import scipy
import numarray
import queue
import thread
import SimpleXMLRPCServer
import csv
import os

#This class implements all functionality our server requires to start a
#simulation processes and communicate with a windows graphical client:

class Server:

#This is an object constructor. It initializes fields that are available in
#the every object's method.

        def __init__(self):
            self.steps = 0
            self.nofs = 0

#This function creates empty, filled with zeros, tables that will contain
#data averaged from simulations that are completed. Tables 'a' and 'b'
#contains the x and y axis, and the array 'res' contains calculated results
#of an exchange-bias and a coercivity:

        def create_tables(self):
          self.res = numarray.zeros((self.nofs,6), numarray.Float32)
          self.a = numarray.zeros((self.nofs,self.steps,3), numarray.Float32)
          self.b = numarray.zeros((self.nofs))

#This function creates the queue of simulations to run, with all the input
#physical parameters and the binary C++ file that contains this simulation.
        #The filed 'self.nofs' contains information about number of
#simulations we want to run, so the 'for' loop creates chosen number of
#simulations to run, and pass it to the 'queue' module, which takes control
#over starting the specified simulation repeat on the next not busy cluster
#node.
        #The 'for' loop also assigns processes identification number, the
#PID, and passes it on as the command line parameter named '--pid'. In this
#way we control a moment when a simulation process was completed.
        #Last line uses the internal python module called 'thread'. It starts
#the new python 'thread' that runs a queue:
```

```python
    def start(self, net_size, steps, hmax, hmin, dlevel, econst, sd, sd1,
pp, nofs, runall, bin_name):
        self.steps = int(steps)
        self.nofs = int(nofs)
        self.create_tables()
        q = []
        for i in range(self.nofs):
            q.append(['./bin/'+bin_name,'--size',net_size,'--
steps',steps,'--hmax',hmax,'--hmin',hmin,'--dlevel',dlevel,'--
econst',econst,'--sd',sd,'--asd',sd1,'--pp',pp,'--pid',str(i)])
        thread.start_new_thread(queue.run,(q,))
        return 0
```

#This is the function that calculates mean value from data returned by
#simulations which were completed. The 't1' array contains a mean value of
#results returned by simulations, the 't2' array contains a standard
#deviation errors of the results (exchange-bias, coercivity); in summary
#there are 4 values. In the last loop they are packed into the 'temp' array
#and returned to a graphical client, which expects them in this particular
#format:

```python
    def res_sum(self):
        has_finish = self.fin_as_far()
        temp = []
        if has_finish > 0:
            t1 = scipy.stats.mean(self.res,0)
            t2 = scipy.stats.stderr(self.res,0)

            for i in range(4):
                temp.append([t1[i],t2[i]])
        return temp
```

#This function determines how many of simulations were completed so far:
```python
    def fin_as_far(self):
        temp = []
        temp = self.getfinished()
        finished = 0
        for i in temp:
            if i == 1:
                finished += 1
        return finished
```

#This is the function called by a simulation process to store calculated
#values of an exchange-bias and a coercivity:
```python
    def store_results(self, pid, results):
        print "RESULTS APP PID:"+str(pid)
        for i in range(6):
            self.res[pid][i] = results[i]
        return 2
```

#This is the function that stores magnetization values (y axis) for
#subsequent values of an external magnetic field; the magnetization values
#are taken from the completed simulation repeats:

```python
    def store_array2(self, pid, array2):
        for i in range(self.steps):
            self.a[pid][i][1] = array2[i]
        return 0
```

#This is the function that stores external magnetic field values (x axis;
#the  values are taken from the completed simulation repeats:

```python
        def store_array1(self, pid, array1):
            for i in range(self.steps):
                self.a[pid][i][0] = array1[i]
            return 0

#This function, called by a simulation process, reports that calculations
#were finished, and all resulting data were sent back to a server:
        def finish(self, pid):
            self.b[pid] = 1
            return 0

#In this line we create an object of the class 'Server'. We need that
#object to register it with the XMLRPC server, to make it's internal
#methods available over the network:

i = Server()

#Last three lines are needed to create XMLRPC server, register it to listen
#on a specific network IP address and a port. This is done in the first
#line, the second one registers the class instance, represented by variable
#'i'. The last line one start internal event loop for XMLRPC server, which
#makes him persistent server, waiting for connections:

server = SimpleXMLRPCServer.SimpleXMLRPCServer(('10.0.0.1', 8000))
server.register_instance(i)
server.serve_forever()
```

## Appendix D. The Python code of the queue.py file which controls repeats of simulated physical phenomenon

```python
#!/usr/bin/env python

import sys, os, time

#This list contains number of nodes available in the cluster. The number
#has to be the same as the open-Mosix nodes (computers) number:

def run(queue):
      nodes = [1, 2, 3, 4, 5, 6, 7, 8, 9, 10, 11, 12, 13, 14, 15]

#This is the array used for keeping pids of currently running processes:

      pids = len(nodes) * [0]

#Keep time when a simulation was started:

      starttime = time.time()

#Maximum number of simultaneous processes running:

      max = len(nodes)

#A number of currently running processes:

      children = 0

#Repeat until there are processes in a queue or just running tasks:

      while queue or children:
            Check if there are less running tasks than maximum possible and

#If there are any tasks to run in a queue:

            if children < max and queue:

#In next few lines we start another process form a 'queue' list and
#remember it's PID (the process identification number):

                  i = pids.index(0)
                  node = nodes[i]

                  pid = os.spawnlp(os.P_NOWAIT, 'mosrun', 'mosrun', '-L'
,'-j%s' % node, queue[0])

#Print out a message on the console that process has been started:

                  print 'Job %s pid %d started on node %d.' % (queue[0],
pid, node)

#Hold the process PID in an list of currently running tasks:

                  pids[i] = pid

#Remove a started task from a process 'queue' list:

                  del queue[0]
```

```python
#Increase the running processes counter:

                children += 1

#If there are no processes in a 'queue' to run or there is no more nodes in
#a cluster left to use

            else:

#then, wait for a processes to be completed, then get the pid and hold it
#in the 'done' varaiable:

                wlist = os.waitpid(-1, 0)
                done = wlist[0]

#If there are PID of completed processes in the 'done' variable that exists
#on the pids list:

            if done in pids:

#then, get the list index where a 'done' PID is:

                i = pids.index(done)

#Print out message that certain process was completed (finished):

                print "Process %d on node %d finished." % (pids[i],
nodes[i])

#Erase the pid number in the PIDs list:

                pids[i] = 0

#Decrease the running processes number:

                children -= 1

#Log the total end time of calculation, and print out the message on a
#console, how long lasted the whole number of processes:

     endtime = time.time()
     minsec = divmod((endtime - starttime), 60)
     print 'Completed in %s min %s sec. ' % (str(int((minsec[0]))),
str(int((minsec[1]))))
```